\documentclass[a4paper,12pt,english]{article}
\usepackage{amsfonts}
\usepackage{graphicx}
\usepackage{amsmath}
\usepackage{color}

\newcommand{\be}{\begin{equation}}
\newcommand{\ee}{\end{equation}}
\begin{document}


\begin{center}
\noindent{\bf{\LARGE{Vacua of Exotic Massive 3D Gravity}}}

\smallskip
\smallskip

\smallskip
\smallskip

\smallskip
\smallskip

\smallskip
\smallskip

\noindent{{Mariano Chernicoff$^a$, Gaston Giribet$^{b}$, Nicol\'as Grandi$^c$,  Julio Oliva$^d$}}

\smallskip

\smallskip

\smallskip

\smallskip
\smallskip

\smallskip
\smallskip

\smallskip

\smallskip

\smallskip

\smallskip

$^a${Departamento de F\'{\i}sica, Facultad de Ciencias, Universidad Nacional Aut\'onoma de M\'exico} \\ {\it A.P. 70-542, CDMX 04510, M\'exico.}

\smallskip

$^b${Center for Cosmology and Particle Physics, New York University}\\ {\it 726 Broadway 10003 New York City, USA.}

\smallskip

$^c${Instituto de F\'isica de La Plata - CONICET \& Departamento de F\'isica - UNLP} \\ {\it C.C. 67, 1900 La Plata, Argentina.}

\smallskip

$^d$ {Departamento de F\'{i}sica, Universidad de Concepci\'on} \\ {\it Casilla 160-C, Concepci\'on, Chile.}

\end{center}

\smallskip

\smallskip

\smallskip

\smallskip

\smallskip

\smallskip

\smallskip

\begin{center}
Abstract
\end{center}
We consider the recently proposed exotic 3D massive gravity. We show that this theory has a rich space of vacua, including asymptotically Anti de-Sitter (AdS) geometries obeying either the standard Brown-Henneaux boundary conditions or the weakened asymptotic behavior of the so-called Log-gravity. Both sectors contain non-Einstein spaces with $SO(2)\times \mathbb{R}$ isometry group, showing that the Birkhoff theorem does not hold all over the parameter space, even if strong AdS boundary conditions are imposed. Some of these geometries correspond to 3D black holes dressed with a Log-gravity graviton. We conjecture that such geometries appear in a curve of the parameter space where the exotic 3D massive gravity on AdS$_3$ is dual to a chiral conformal field theory. The theory also contains other interesting vacua, including different families of non-AdS black holes.



\newpage

\section{Introduction}

Three-dimensional (3D) gravity shares with its four-dimensional analog many interesting qualitative features; the most salient one being the existence of black holes \cite{BTZ}. At the same time, it turns out to be much more accessible, what makes it a perfect toy model to explore theoretical aspects of gravity to which, otherwise, we would not have access. For example, it permits to give a microscopic description of non-supersymmetric black holes entropy \cite{Strominger}, to sum over geometries and discuss the saddle points content of the quantum gravity partition function \cite{MaloneyWitten}, or even to investigate the consistency of a holographic dual for a pure gravity theory \cite{Witten}.

One of the simplifications that Einstein theory in 3D presents with respect to 4D is the absence of local degrees of freedom, what reduces the content of the theory to defects \cite{3D, 3D2}, black holes \cite{BTZ, BTZ2}, and boundary gravitons \cite{BH}, but with no presence of propagating modes. The theory, however, does acquire propagating modes when one deforms it by adding mass to the graviton. Indeed, in 3D there exist consistent ways of giving mass to the graviton. One such massive deformation is the well-known topologically massive gravity (TMG) \cite{TMG}, which is defined by augmenting the Einstein-Hilbert action with a Chern-Simons term for the affine connection. The interest of such model has been revived some years ago within the context of the AdS/CFT correspondence, specially in relation to the so-called chiral gravity \cite{Chiral}; see also \cite{GJ, GKP, Deser1, Deser2, Log}. 

Another consistent way of giving mass to the graviton in 3D is the so-called new massive gravity (NMG), which is defined by adding to the action a particular combination of higher-curvature terms that suffices to decouple the ghostly scalar mode \cite{NMG}. Both TMG and NMG, however, present a consistency problem when discussed in the context of AdS/CFT. This problem is known as the bulk-boundary clash, and basically means that there is no way of achieving a unitary theory in bulk and in the boundary simultaneously. Among the attempts to solve this problem, a new and arguably simpler massive deformation of 3D Einstein gravity, known as minimal massive gravity (MMG), was proposed \cite{MMG}. This theory is defined by supplementing TMG equations of motion with a second-order rank-2 tensor. In 3D dimensions, this implies that such a tensor, being of second order in the metric, cannot follow from a variational principle in the metric formalism, what is typically problematic for the Bianchi identities. However, MMG manages to circumvent this obstruction in a very interesting way, that is, even though the Bianchi identities are not identically satisfied by the tensors involved in the equations of motion, they do hold on-shell. This way of solving the consistency conditions is usually referred to as the third way phenomenon \cite{thirdway}. 

More recently, a fourth-order massive deformation of 3D gravity exhibiting the same kind of third way phenomenon has been proposed \cite{Exotic}. This has been dubbed exotic massive 3D gravity (EMG). It is defined by the following equations of motion
\begin{equation}
R_{\mu \nu }-\frac {1}{2} Rg_{\mu \nu }+\Lambda g_{\mu \nu } + \frac{1}{\mu }C_{\mu \nu} - \frac{1}{m^2} H_{\mu \nu} + \frac{1}{m^4}L_{\mu \nu } =0\label{Prima}
\end{equation}
where 
\begin{equation}
C_{\mu \nu} = \frac 12 \epsilon_{\mu}^{\ \alpha \beta}\nabla_{\alpha} R_{\beta \nu } +  \frac 12 \epsilon_{\nu}^{\ \alpha \beta}\nabla_{\alpha} R_{\beta \mu }  \ , \ \ \ 
H_{\mu \nu } = \epsilon_{\mu }^{\ \alpha \beta}\nabla_{\alpha}C_{\nu \beta } \ , \ \ \ L_{\mu \nu }=\frac{1}{2} \epsilon_{\mu }^{\ \alpha \beta} \epsilon_{\nu }^{\ \gamma \sigma} C_{\alpha \gamma} C_{\beta \sigma }.\label{Segunda}
\end{equation}
$C_{\mu \nu}$ is the Cotton tensor, which also appears in TMG. In fact, TMG corresponds to the limit $m\to \infty $ of the theory above. 

The covariant divergence of the tensors $H_{\mu\nu}$ and $L_{\mu\nu}$ does not vanish identically, but the following identities hold
\begin{equation}
\epsilon_{\mu}^{\ \alpha \beta }C_{\alpha}^{\sigma}(R_{\beta\sigma}-\frac 12 Rg_{\beta\sigma}-\frac{1}{m^2} H_{\beta\sigma} )=\epsilon_{\mu}^{\ \alpha \beta }(-\Lambda C_{\alpha }^{\sigma}g_{\beta\sigma} -\frac{1}{m^4} C_{\alpha }^{\sigma}L_{\beta\sigma})=0\label{This}
\end{equation}
and, in virtue of (\ref{This}), one finds that the Bianchi identities can be satisfied on-shell without imposing incompatible constraints. 

Both tensors $H_{\mu \nu }$ and $L_{\mu \nu }$ are defined in terms of the Cotton tensor, and so they vanish for conformally flat metrics. In 3D, all Einstein manifolds are locally equivalent to maximally symmetric spaces, and hence conformally flat. Therefore, general relativity (GR) appears as a subsector of the space of solutions of (\ref{Prima})-(\ref{Segunda}). Here, we will be concerned with solutions to these equations that are not Einstein spaces. This will enable us to investigate different aspects of the theory, such as the mass of the excitations around maximally symmetric spaces like AdS, the black holes content of the theory and their (non-)uniqueness, and the existence of other backgrounds and asymptotic conditions that might be of interest for physics applications.

In section 2, we will consider the non-linear regime of the theory. We will study gravitational wave solutions in AdS space and the effective mass of such solutions, which is found to agree with the analysis of linearized modes found in \cite{Exotic}. In section 3, we focus on asymptotically AdS non-Einstein spaces that are solutions of the theory at a special point of the parameter space that can be thought of as the analog of the chiral point of TMG. We discuss solutions with both strong and weak falling-off behavior in AdS. In section 4, we discuss other vacua of the theory, including Warped AdS black holes, Lifshitz black holes, among others.  

\section{The non-linear theory}

In this section and the next one we will mainly focus on solutions to (\ref{Prima})-(\ref{Segunda}) when the following relation among the parameters is satisfied
\begin{equation}
\mu =\frac{m^2\ell}{1-m^2\ell^2} \label{chiral}\ ,
\end{equation}
where $\ell^2=-1/\Lambda$ (hereafter we will consider $\ell=1$). We will refer to (\ref{chiral}) as the {\it critical point} (or, more precisely, the chiral {\it curve}). When this relation is obeyed, the theory exhibits quite special features: The linear excitations around AdS$_3$ become massless \cite{Exotic} and low-decaying logarithmic modes appear. This can also be observed at non-linear level by studying gravitational wave solutions on AdS$_3$ of the type analyzed in \cite{EloyMokhtar, EloyMokhtar2, Bending}. For generic $\mu$ and $m$, one considers the ansatz
\begin{equation}
ds^2 = -r^2 N_w(u,r) du^2-2r^2dudv+\frac{dr^2}{r^2}\ ,\label{wave}
\end{equation}
and take $v\in \mathbb{R}$, $u\in \mathbb{R}$, $r\in \mathbb{R}_{\geq 0}$. This ansatz represents a gravitational wave in AdS$_3$, where $v$ and $u$ are two null directions, analogously to a $pp$-wave. Function $N_w(u,r)$ describes the profile of the wave. In the case $N_w=const$ the solution is locally equivalent to AdS$_3$. 

Ansatz (\ref{wave}) solves the equations of motion (\ref{Prima})-(\ref{Segunda}) for
\begin{equation}
N_w(u,r) = c_0(u)+c_2(u)r^{-2}+c_+(u)r^{\alpha_{+}}+c_-(u)r^{\alpha_{-}} \ , \label{H}
\end{equation}
with 
\begin{equation}
\alpha_{\pm} = \frac{m^2+ 2\mu\pm\sqrt{m^4+4m^2\mu^2}}{2\mu}\label{HM}
\end{equation}
provided $\mu \neq ({m^2\ell})/({1-m^2\ell^2})$. Here, $c_i(u)$ ($i=0,2,\pm $) are functions of the null coordinate $u$. When $c_{+}=c_{-}=0$ one obtains the GR solutions, while the modes with Yukawa decaying, $c_{\pm}\neq 0$, represent the massive gravitons of the theory. If one considers the particular solution $N_w^{\pm }=c_{\pm}(u)r^{\alpha_{\pm}}$, i.e. setting $c_0=c_2=c_{\mp }=0$ in (\ref{H}), then such solution satisfies the wave equation
\begin{equation}
\Big(\Box +M^2_{\pm}+\Lambda\Big) N_w^{\pm }(u,r)=0,
\end{equation}
with 
\begin{equation}
M_{\pm } = \frac{m}{2\mu }\Big( \sqrt{m^2+4\mu^2}\pm m\Big),
\end{equation}
where $\Box$ is the D'Alambertian operator associated to the metric (\ref{wave}). This value of $M_{\pm }$ exactly agrees with the mass of the modes obtained by the linearized analysis of \cite{Exotic}. Here, we have  obtained the same result but from the non-linear analysis. 

In contrast, at the critical point $\mu = ({m^2\ell})/({1-m^2\ell^2})$ the solution for $N_w$ takes a different form; namely
\begin{equation}
H(u,r) = c_0(u)+c_2(u)r^{-2}+c_+(u)\log (r)+c_-(u)r^{-1-m^2} 
\end{equation}
where we explicitly see the presence of the low-decaying modes. In the next section we will discuss these logarithmically decaying solutions in more detail. 

\section{Non-Einstein geometries in AdS$_3$}

\subsection{Strong boundary conditions}

 We expect the theory on the {\it critical curve} (\ref{chiral}) to be dual to a chiral CFT$_2$ with central charges $c_L=0$, $ c_R={3}/|{G\mu}| $. However, this value for $c_R$, i.e. twice that of conformal gravity, assumes that the action of EMG do not contribute to the diffeomorphism anomaly. For $\mu =\infty$ the equations of motion (\ref{Prima})-(\ref{Segunda}) are parity-even, but the Chern-Simons type action of the theory does violate parity; see \cite{Exotic} for details.

Equations (\ref{Prima})-(\ref{Segunda}) admits, of course, locally AdS$_3$ geometries as solutions, and in particular those that are asymptotically AdS$_3$ in the Brown-Henneaux sense \cite{BH}. These are solutions that in a given system of coordinates behave like
\begin{eqnarray}
g_{tt} &\simeq& r^2 + {\mathcal O}(1) \ , \ \ g_{t\phi } \simeq {\mathcal O}(1) \ , \ \  g_{\phi \phi } \simeq r^2 + {\mathcal O}(1) \ , \label{AS}\\
g_{rr} &\simeq& r^{-2}+ {\mathcal O}(r^{-4}) \ , \ \ g_{tr} \simeq {\mathcal O}(r^{-4}) \ , \ \ g_{r\phi } \simeq {\mathcal O}(r^{-4}) \ ,\label{BS}
\end{eqnarray}
where ${\mathcal O}(r^n)$ stand for terms of order $r^n$ or subleading. We take $t\in \mathbb{R}$, $r\in \mathbb{R}_{\geq 0}$, $\phi \in [0,2\pi ]$. Solutions (\ref{wave}), for example, obey these conditions provided $\alpha_{\pm }\leq 0$.

The point we want to make first is that Einstein spaces are not the only ones that obey the above boundary conditions. There exist, in addition, non-Einstein solutions that also asymptote to AdS$_3$ in the Brown-Henneaux sense (\ref{AS})-(\ref{BS}). One such geometry is given by 
\begin{equation}
ds^{2}=-r^{2}dt^{2}+\frac{dr^{2}}{r^{2}}+r^{2}d\phi ^{2}+N_{\gamma
}(t,r)(dt+d\phi )^{2}  \label{A}
\end{equation}%
with 
\begin{equation}
N_{\gamma }(t,r)=\left( \beta (t-t_0)+\frac{\gamma }{r^{4}}\right) .
\label{Ng}
\end{equation}%
It can be verified that this \textit{ansatz} solves the field equations if
\begin{equation}
 \beta^2 (m^4 + 5m^2-2) +96  \gamma ( m^4 +5 m^2)  = 0 ,\label{Lanueva}
\end{equation}%
with $t_0$ arbitrary. Notice that the solution exhibits time translation symmetry even though (\ref{Ng}) explicitly depends on $t$. Despite being non-Einstein spaces, solutions (\ref{A})-(\ref{Ng}) do obey the Brown-Henneaux asymptotic conditions (\ref{AS})-(\ref{BS}). In the limit $m\to \infty$, one obtains $\mu =-1$ and $\beta^2=-96\gamma$, which is the result for TMG \cite{Sophie}. Metric (\ref{A})-(\ref{Ng}) exhibits closed timelike curves due to the fact that function $N_{\gamma }(t,r)$ is unbounded. This, however, does not affect the signature of the metric. In the case $\beta=0$, where $N_{\gamma }=0$, the metric reduces to that of the massless BTZ. For $\beta \neq 0$, the metric is not conformally flat. 

This type of solution is important for several reasons. In particular, it manifestly shows that Birkhoff theorem does not hold in Exotic Massive Gravity, at least at the critical point (\ref{chiral}). And this is the case even if the strong boundary conditions are considered. This is relevant for the discussion of the bestiary of geometries that contribute to the partition function of the quantum theory, cf. \cite{Log}. 

\subsection{Weak boundary conditions}

The theory also admits solutions that, while not obeying the Brown-Henneaux boundary conditions, do respect the weakened AdS$_3$ asymptotic behavior proposed in \cite{GJ}, which leads to the definition of the so-called Log-gravity \cite{Log}; namely 
\begin{eqnarray}
g_{tt} &\simeq& r^2 + {\mathcal O}(\log r) \ , \ \ g_{t\phi } \simeq {\mathcal O}(\log r) \ , \ \  g_{\phi \phi } \simeq r^2 + {\mathcal O}(\log r ) \ , \label {FG}\\
g_{rr} &\simeq& r^{-2}+ {\mathcal O}(r^{-4}) \ , \ \ g_{tr} \simeq {\mathcal O}(1) \ , \ \ g_{r\phi } \simeq {\mathcal O}(1) \ , \label {GF}
\end{eqnarray}
cf. (\ref{AS})-(\ref{BS}). An example of such a solution is given by considering the extremal Ba\~{n}ados-Teitelboim-Zanelli black hole \cite{BTZ}
\begin{equation}
ds^{2}=-N^{2}(r)dt^{2}+\frac{dr^{2}}{N^{2}(r)}+r^{2}(N_{\phi }(r)dt+d\phi
)^{2}  \label{UNO}
\end{equation}%
with
\begin{equation}
N^{2}(r)=\frac{(r^2-r^2_H)^2}{r^2}\ ,\ \ \ \ \ N_{\phi }(r)=\frac{r_H^2}{r^{2}},\label{UNA}
\end{equation}%
at the critical point (\ref{chiral}), and perturbing it by adding to the metric a term \cite{GAY}
\begin{equation}
M\log (r^{2}-r^2_H)\ (dt+d\phi )^{2} \ ,    \label{DOS}
\end{equation}
where $M$ and $r_H$ are arbitrary real constants. This solution is also solution in the limit $m\to \infty$ \cite{GAY, Garcia}. In that case, the mass has been shown to be proportional to $M$, which is not the mass of the BTZ to which the solution reduces in the limit $M\to 0$. The extremal black hole $M= 0$ has the event horizon at $r=r_H$. There, the logarithm in (\ref{DOS}) for the solution with $M\neq 0$ diverges. Nevertheless, the curvature scalars of the metric remain finite all over the space. The metric can be continued to the region $r<r_H$ by changing the sign in the argument of the logarithm in (\ref{DOS}).

In brief, metric (\ref{UNO})-(\ref{DOS}) represents a black hole dressed with a Log-gravity graviton and realizes the boundary conditions (\ref{FG})-(\ref{GF}) at the non-linear level. 

\section{Other vacua}

\subsection{Warped black holes}

Besides those of GR and the ones discussed above, theory (\ref{Prima})-(\ref{Segunda}) admits other interesting type of solutions, which appear on other curves of the parameter space ($\mu,m$). In particular, it admits Warped Anti-de Sitter (WAdS) spaces
\begin{equation}
ds^2 = \frac{\ell^2}{\nu^2+3} \Big( -\cosh^2 r \ dt^2 +dr^2 +\frac{4\nu^2}{\nu^2+3} (d\phi +\sinh r \ dt)^2\Big) \label{WAdS}
\end{equation}
with the parameters $\nu $ and $\ell^2$ being given in terms of the couplings $m$, $\mu$, $\Lambda$ as follows
\begin{equation}
\mu = \frac{\ell^3\nu m^4}{3(\nu^4-\nu^2(1+\ell^2 m^2)+\ell^4m^4/9)}  \ , \ \ \ \ \nu^6-2\nu^4+\nu^2+m^4\ell^4 (1+\Lambda \ell^2)/9=0.\label{para}
\end{equation}

These metrics represent squashed ($\nu <1$) or stretched ($\nu >1$) deformations of AdS$_3$ space. The dimensionless parameter $\nu $ controls the squashing effect (the shape), while the dimensionful parameter $\ell^2$ gives the curvature radius (the size). This metric can be thought of as a double Wick rotation of a 3D section of G\"{o}del solution of 4D cosmological Einstein equations. The value $\nu =1$ in (\ref{WAdS}) corresponds to undeformed AdS$_3$ space written as a Hopf fibration of AdS$_2$. There is also a limit in which metric (\ref{WAdS}) describes a AdS$_2\times S^1$ space, but since both $H_{\mu\nu}$ and $L_{\mu\nu}$ are transparent to conformally flat solutions, these AdS$_2$ vacua only appear in the conformal limit $\mu \to 0$.

There exist black holes that asymptote to WAdS$_3$ space (\ref{WAdS}) at large distance \cite{Clement}. Their 2-parameters metric, in a convenient system of coordinates, can be found in \cite{Anninos}. It turns out that these black holes are, in addition, locally equivalent to WAdS$_3$ \cite{Anninos}, and therefore are also solutions of the EMG theory when the relation (\ref{para}) is satisfied.

\subsection{Lifshitz black holes}

Besides AdS black hole and WAdS black holes, equations of motion (\ref{Prima})-(\ref{Segunda}) in the limit $\mu \to \infty$ admit as solutions black hole geometries that asymptote to spaces with an anisotropic scale invariance. These are the so-called Lifshitz black holes; see \cite{Lifshitz} and references therein and thereof. The metric of these black holes have the form 
\begin{equation}
ds^2 = -\frac{r^{2z}}{\ell^{2z}} \Big( 1-\frac{r^2_H}{r^2}\Big) dt^2 +\frac{\ell^2 }{r^2} \Big( 1-\frac{r^2_H}{r^2}\Big)^{-1}dr^2 + r^2 d\varphi^2 \label{Lifshitz}
\end{equation}
with $r^2_H$ being an arbitrary integration constant that gives the radial position of the horizon. We take $t\in \mathbb{R}$, $r\in \mathbb{R}_{\geq 0}$, $\varphi \in \mathbb{R}$. Metric (\ref{Lifshitz}) solves the equations of motion of EMG provided that the dynamical exponent and the curvature radius satisfy the relations
\begin{equation}
z= m^2\ell^2 \ , \ \ \ \ \Lambda = -m^2 .
\end{equation}
Metrics (\ref{Lifshitz}), when $r_H\neq 0$ have non-constant curvature scalars, unlike the geometries studied above. 

This gives a whole family of asymptotically Lifshitz black holes with arbitrary value of the dynamical exponent $z$, which is set by the mass coupling $m$. At the chiral point (\ref{chiral}), since $\mu =\infty$ and $m^2\ell^2=1$, one finds that the solution (\ref{Lifshitz}) coincides with the BTZ black hole, $z=1$ with $\Lambda=-1/\ell^2$. The solution with $z=3$ is also present in the case of NMG. The case $z=2$ is particularly interesting in the holographic description of condensed matter systems as it describes finite-temperature Lifshitz fixed points in 1+1 dimensions.    

\subsection{Euclidean vacua}

The theory also contains other solutions at different points of the parameter space. Let us show another two examples, which correspond to particular cases of Thurston geometries \cite{Thurston}. The latter are relevant in the study of the 3D uniformization problem. One such geometry is given by the Sol (for solvable) metric
\begin{equation}
ds^2 = dr^2 + e^{2r} dx^2 + e^{-2r} dy^2 
\end{equation}
which is an Euclidean metric that solves the equations (\ref{Prima})-(\ref{Segunda}) for $\mu=\infty $ if either the condition $m^2+1=1-\Lambda=0$ or the condition $m^2+2=\Lambda =0$ holds. A related example is given by the Nil (for nilpotent) metric
\begin{equation}
ds^2 = dx^2 + dy^2 +(dr-xdy)^2
\end{equation}
which also solves (\ref{Prima})-(\ref{Segunda}) for $\mu=\infty $ if $8m^2=-9\pm \sqrt{33}$ and $8\Lambda = 17/3\pm \sqrt{33}$ hold. This illustrates the variety of geometries that EMG contains in its space of solutions. 

\section{Final remarks}

We have shown that the exotic theory of massive 3D gravity proposed in \cite{Exotic} has a rich space of vacua. We have provided several examples of such geometries, including gravitational waves, asymptotically AdS$_3$ non-Einstein spaces obeying either strict or relaxed boundary conditions, Lifshitz black holes with arbitrary value of dynamical exponent, Warped-AdS$_3$ spaces including Warped-AdS$_3$ black holes and their locally equivalent G\"{o}del type solutions, among others. Some of these solutions, like the $1\neq z\neq 3$ Lifshitz black holes, do not exist neither in TMG nor NMG. In particular, we have shown that non-Einstein spaces with $SO(2)\times \mathbb{R}$ isometry group exist at the critical point (\ref{chiral}). This implies that the Birkhoff theorem does not hold in EMG, even when the Brown-Henneaux boundary conditions are imposed.

\section*{Acknowledgments}

The work of G.G.  is supported in part by the NSF through grant PHY-1214302. M.C. is partially supported by Mexico's National Council of Science and Technology (CONACyT) grant 238734 and DGAPA-UNAM grant IN113618. This work was also partially supported by CONICYT grant PAI80160018 and Newton-Picarte Grants DPI20140053.

\end{document}